# Biosorption of Cr(VI)_ and Cr(III)_Arthrobacter species

*E.Gelagutashvili*, E.Ginturi* D.Pataraia**, M.Gurielidze***

**Iv. Javakhishvili Tbilisi State University*
*E. Andronikashvili Institute of Physics*
*0177, 6, Tamarashvili St.,*
***Durmishidze Institute of Biochemistry and Biotechnology,*
*D.Agmashenebeli Kheivani,10 km,0159,*
*Tbilisi, Georgia*

## Abstract

The biosorption of Cr(VI)_ and Cr(III)_ *Arthrobacter* species (*Arthrobacter globiformis* and *Arthrobacter oxidas*) was studied simultaneous application dialysis and atomic absorption analysis. Also biosorption of Cr(VI) in the presence of Zn(II) during growth of *Arthrobacter* species and Cr(III) in the presence of Mn(II) were discussed.

Comparative Cr(VI)_ and Cr(III)_ *Arthrobacter* species shown, that Cr(III) was more effectively adsorbed by both bacterium than Cr(VI). The adsorption capacity is the same for both the Chromium-*Arthrobacter* systems. The biosorption constants for Cr(III) is higher than for Cr(VI) 5.7-5.9- fold for both species.

Is was shown significant difference between the binding constants for Cr(VI) –*A.oxidas* and Cr(VI)-*A.globiformis*.

Comparative Freundlich biosorption characteristics Cr(VI) *Arthrobacter* species of living and dry cells shown, that capacity(*n*) is in both cases the same(1.25,1.35). Dry cells have larger biosorption constant for both species (*K*)( 4.6x $10^{-4}$, 3.4x$10^{-4}$, than living cells(1.0x $10^{-4}$ ,1.36x $10^{-4}$ ).

Biosorption characteristics (K) and (*n*) for *A. oxidas* are without Mn(II) and in the presence of Mn(II) 2.6 x $10^{-4}$ (*K*), 1.37 (n) and 2.4 x $10^{-4}$ (*K*) '1.41 (*n)* respectively; for *A. globiformis* without Mn(II) and in the presence of Mn(II) 2.0 x $10^{-4}$ (*K*), 1.23 (*n*) and 1.9 x $10^{-4}$ (*K*), 1.47 (*n*) respectively. Thus, Biosorption characteristics did not change in the presence of Mn(II) ions..

It was shown, that bioavailability increases in the presence of Zn ions in both cases. (for Cr(VI)-*Arthrobacter globiformis* and for Cr(VI)- *Arthrobacter oxidas*). Biosorption characteristics (*K*) and (*n*) for A. oxides in the absence and in the presence of Zn(II) are 4.6 x $10^{-4}$ (*K*), 1.25 (*n*) and 6.6 x $10^{-4}$ '1.08 (*n*) respectively, for *A. globiformis* without Zn(II) and in the presence of Zn(II) 3.4 x $10^{-4}$ (*K*), 1.35 (*n*) and 8.1 x $10^{-4}$ (*K*), 1.19 (*n*) respectively; n values did not change. But for Cr(VI)-*Arthrobacter globiformis* increase is more significant.

## Introduction

Chromium exists in the environment mainly as Cr(III) and Cr(VI) species. The interest in Cr is governed by the fact that its toxicity depends critically on its oxidation state. While Cr(III) is considered essential lipid and protein metabolism, Cr(VI) is known to be toxic to humans[1]. Gram-positive Arthrobacter species bacteria can reduce Cr(VI) to Cr(III) under aerobic growth and there is a large interest in Cr-reducing bacteria. The exact mechanism by which microorganisms take up the metal is relatively unclear.

Metal sorption performance depends on some external factors such as pH, other ions in solution, which may be in competition ets. Several aerobic and anaerobic Cr(VI) reducers are known with some being able to use organic contaminants as electron donors for Cr(VI) reduction[2].

The inhibitive effect of FeS on Cr(III) oxidation by biogenic Mn-oxides that were produced in the culture of a known species of Mn(II) oxidizers, *Pseudomonas putida*. In soils containing manganese oxides, the immobilized form of chromium Cr(III) could potentially be reoxidized.[3,4]. Arthrobacter species strain FR-3, isolated from sediments of a swamp, produced a novel serine type oxidase. The purified free sulfide oxidase activity was completely inhibited by Co(II) and Zn(II) [5]. In[6] was shown, that Chromium reductase activity of Arthrobacter rhombi-RE strain was associated with the cell-free extract and the contribution of extracellular enzymes to Cr(VI) reduction was negligible. Ca(II) enhanced the enzyme activity, while Hg(II), Cd(II) and Zn(II) inhibited the enzyme activity.

Environmental systems are always dynamic and often far from equilibrium. In spite of this, the Biotic Ligand Model assumes that the metal of interest and its complexes are chemical equilibrium with each other and with sensitive sites on the biological surface [7]. Constants for the interaction of the metal with the biological surface have been estimated by measuring metal internalization fluxes, metal loading, and metal toxicity [1]. To develop an efficient biosorbent and its reuse by subsequent desorption processes, knowledge of the mechanism of metal binding is thus very important.

The biosorption of Cr(VI) and Cr(III) on *Arthrobacter* species was examined in this study simultaneous application dialysis and atomic absorption analysis. Also biosorption of Cr(VI) in the presence of Zn(II) during growth of *Arthrobacter* species and Cr(III) in the presence of Mn(II) were studied.

## Materials and Methods

The other reagents were used: $K_2CrO_4$, $CrCl_3$, $ZnSO_4$, $MnCl_2x4H_20$ (Analytical grade). *Arthrobacter* bacterials were cultivated in the nutrient medium without co-cations and loaded with concentration of Zn (50mg/l)(in the case of Cr(VI) and Mn(50mg/l)( (in the case of Cr(III) . Cells were centrifuged at 12000 rpm for 10 min and washed three times with phosphate buffer (pH 7.1). The centrifuged cells were dried without the supernatant solution until constant weight. After solidification( dehydrated) of cells (dry weight) solutions for dialysis were prepared by dissolving in phosphate buffer. This buffer was used in all experiments. A known quantity of dried bacterium suspension was contacted with solution containing a known concentration of metal ion. For biosorption isotherm studies, the dry cell weight was kept constant (1 mg/ml), while the initial chromium concentration in each sample was varied in the interval ($10^{-3}$ -$10^{-6}$ M). All experiments were carried out at ambient temperature. Metal was separated from the biomass with the membrane, which thickeness was 30 μg Visking (serva) and analyzed by an atomic absorption spectrophotometer. Dialysis carried out during 72 h.

Data Analysis. The isotherm data were characterized by the Freundlich [8] equation

$$C_b = K C_t^{1/n}$$

where $C_b$ is metal concentration adsorbed on either live or dried cells of *Arthrobacter* species in $mgg^{-1}$ dry weight. $C_t$ is the equilibrium concentration of metal ($mgl^{-1}$) in the solution. $K$ is an empirical constant that provides an indication $logC_b$ as a function of $log\ C_t$ of the adsorption capacity of either live or dry cells, $1/n$ is an empirical constant that provides an indication of the intensity of adsorption. The adsorption isotherms were obtained by plotting $logC_b$ as a function of $log\ C_t$.

## Results and Discussions

Metal removal by living or dry cells of bacterium species (*Arthrobacter globiformis* and *Arthrobacter oxidas*) was studied as a function of metal concentration. The linearized adsorption isotherms of Cr ion in anion and cation forms for two kinds of *Arthrobacter* at room temperature are shown in Fig.1-4 by fitting experimental points. Freundlich parameters evaluated from the isotherms with the correlation coefficients are given in table 1.

In fig. 1 are presented Biosorption Isotherms for Cr(VI)-*Arthrobacter globiformis* and Cr(VI)-*Arthrobacter oxidas* for dry (A,B) and living cells (C,D). As shown in fig.1 the adsorption of Cr(VI) in both cases of *Arthrobacter* species to living and dry cells of *Arthrobacter* species was dependent on their concentrations, and thus fitted the Freundlich adsorption isotherm. The Freundlich adsorption model were used for the mathematical description of the biosorption of Cr *Arthrobacter* species. The correlations between experimental data and the theoretical equation were extremely good, with $R^2$ above 0.90(table 1) for all the cases. The higher correlation coefficient show that the Freundlich model is very suitable for describing the biosorption equilibrium of Chromium by the *Arthrobacter* species in the studied concentration range. (The constants determined in a given concentration range will not necessarity be the same as constants determined in another concentration range, because each determination will have its own detection window [9].

The adsorption yields determined for each *Arthrobacter* were compared in table 1.The data in table 1 show a significant difference between the binding constants for Cr(VI) –*A.oxidas* and Cr(VI)-*A.globiformis*.( Biosorption constants for *Arthrobacter oxidas* and *Arthrobacter globiformis* are 4.6x $10^{-4}$, 3.4x$10^{-4}$ respectively). Decrease in bioavailability has been observed experimentally for Cr(VI)-*Arthrobacter globiformis* as compared with *Arthrobacter oxidas*. It is in good agreement, with literatute data by which, there is a large difference in the efficiency of adsorption in each species of microorganisms, since the sorption depends on the nature and the composition of the cell wall [10]. Metal concentrations sorbed by bacterium and those in solution at equilibrium obeyed the Freundlich equation, suggesting the presence of heterogeneous sorption sites on bacterium surfaces. On the other hand, Gram-positive bacteria have a greater sorptive capacity due to their thicker layer of peptidoglycan which contains numerous sorptive sites[11]

Biosorption involves a combination of active and passive transport mechanisms starting with the diffusion of the metal ion to the surface of the microbial cell. Once the metal ion has diffused to the cell surface, it will bind to sites on the cell surface which exhibit some chemical affinity for the metal. It is known, that plasma membrane is the primary site of interaction of trace metal with living organisms. All biological surfaces contain multiple sites including biotic ligands, transport sites, or specific sites and non-specific active sites that are unlikely to participate in the internalization process, including cell wall polysaccharides, also proteins, and lipids, which act as a basic binding site of heavy metals. Typical biosorption includes two phases. The first one is associated with the external cell surface, which was discussed and the second one is an intracellular biosorption, depending on the cellular metabolism [12]. Functional groups within the wall provide the amino, carboxylic, sulfydryl, phosphate, and thiol groups that can bind metals [13**]. **Once inside the cell, chromium is reduced by cellular components such as glutathione, cysteine, ascorbate. The bacterial ability for Cr(VI) reduction does not require high energy input nor toxic chemical reagents and it allows the use of native, non hazardous strains[14]. It was shown, that the carboxyl groups were the main binding site in the cell wall of gram positive bacteria [15]. Such bond formation could be accompanied by displacement of protons and is dependent in part on the extent of protonation, which is determined by the pH [16].

Proceed from the assumption, may be speculate, that first binding sites for Cr(VI) on the surface of *Arthrobacter* species are carboxyl groups. Our results indicated that Cr(VI) sorption is depended of species of bacterial Arthrobacter. Differences between Arthrobacter species in metal ion binding may be due to the properties of the metal sorbates and the properties of bacterium (functional groups, structure and surface area, depending on the species). Functional groups, such as amino, carboxylic, suphydryl, phosphate and thiole groups, differ in their affinity and specifity for metal binding. n values which reflects the intensity of sorption presents the same trend but, as seen from table 1 for both Arthrobacter species *n* values are not significantly different and their sorption intensity indicator are generally small(1.08-1.47).

Comparative Freundlich biosorption characteristics Cr(VI) *Arthrobacter* species of living and dry cells shown (table 1) , that n values are in both cases the same(1.25,1.35). Dry cells have larger biosorption constant for both species (K)( 4.6x $10^{-4}$, 3.4x$10^{-4}$, than living cells(1.0x $10^{-4}$ ,1.36x $10^{-4}$ This may confirm the hypothasis that metal sorption by this bacterium is independent of the metabolic state of the organism [17].

Comparative Cr(VI)-and Cr(III)- *Arthrobacter* species (fig1 and fig.3, table 1) shown, that Cr(III) was more effectively adsorbed by both bacterium than Cr(VI). The adsorption capacity is the same for both the Chromium-*Arthrobacter* systems. The biosorption constants for Cr(III) is higher than for Cr(VI) 5.65-5.88 fold for both species (table1). Cr(VI) is one of the more stable oxidation states, the others being chromium(II), chromium (III). Cr (VI) can be reduced to Cr(III) by the biomass through two different mechanisms[18]. The first mechanism, Cr (VI) is directly reduced to Cr(III) in the aqueos phase by contact with the electron-donor groups of the biomass. The second mechanism consists of three steps. The binding of anionic Cr(VI) ion species to the positively charged groups present on the biomass surface, the reduction of Cr(VI) to Cr(III) by adjacent electron-donor groups and the release of the Cr(III) ions into the aqueous phase due to electronic repulsion between the positively charged groups and the Cr(III) ions.. The ,,uptake-reduction'' model for chromium(VI) carcinogenicity is, that tetrahedral chromate is actively transported across the cell membrane via mechanisms in place for analogous such as sulfate, $SO_4^{2-}$. Chromium (III) is not actively transported across the cell membrane to lack of transport mechanisms for these octahedral complexes. Thus, Cr (VI) may be adsorbed to bacterium a much lower degree than Cr(III), which was shown in our results.

It is seen from fig.2 that bioavailability increases in the presence of Zn ions in both cases. (for Cr(VI)-*Arthrobacter globiformis* and for Cr(VI)- *Arthrobacter oxidas*). Biosorption characteristics (K) and (n) for A. oxides in the absence and in the presence of Zn(II) are 4.6 x $10^{-4}$ (K), 1.25 (n) and 6.6 x $10^{-4}$ '1.08 (n) respectively, for *A. globiformis* without Zn(II) and in the presence of Zn(II) 3.4 x $10^{-4}$ (K), 1.35 (n) and 8.1 x $10^{-4}$ (K), 1.19 (n) respectively, n values did not change. But for Cr(VI)-*Arthrobacter globiformis* increase is more significant. The binding data are in good agreement with literature data, by which biological ligands are generally polyfunctional and polyelectrolytic, with an average pK value between 4.0 and 6.0 [2, 19]. The presence of other cations (Zn(II) increased the uptake of the target cations by bacterium. Such effect from other cation (Zn(II) suggest that at least ion exchange is one of the mechanisms responsible for metal uptake by such Arthrobacter species. This has implications for the selection of *Arthrobacter* species for indrustrial applications.

Different species of bacterium displayed a different sorptive relationship. Biosorption is often followed by a slower metal binding process in which additional metal ion is bound, often irreversibly. This slow phase of metal uptake can be due to a number of mechanisms, including covalent bonding, crystallization on the cell surface or, most often, diffusion into the cell interior and binding to proteins and other intercellular sites [20]. Biosorption may be associated not only to physico-chemical interactions between the metal and the cell wall, but also with other mechanisms, such as the microprecipitation of the metal[21]or the metal penetration through the cell wall[22]

Biosorption characteristics (K) and (n) for *A. oxidas* are without Mn(II) and in the presence of Mn(II) 2.6 x $10^{-4}$ (*K*), 1.37 (*n*) and 2.4 x $10^{-4}$ (*K*), 1.41 (*n*) respectively, for *A. globiformis* without Mn(II) and in the presence of Mn(II) 2.0 x $10^{-4}$ (*K*), 1.23 (*n*) and 1.9 x $10^{-4}$ (*K*), 1.47 (*n*) respectively. Thus, Biosorption characteristics did not change in the presence of Mn(II) ions.. This means, that Mn(II) did not significantly affect the biosorption of Cr (III) ion-*Arthrobacter* species i.e. Mn(II) essentially did not displace Cr(III) from bacteria. This fact leads us to speculate that primary binding site for Cr(III) is different than the binding site for Mn(II). The distorted octahedral coordination sphere proposed for Cr(III) and strong tendency to coordinate donor atoms equatorially may be responsible for the specific interaction with *Arthrobacter* species.

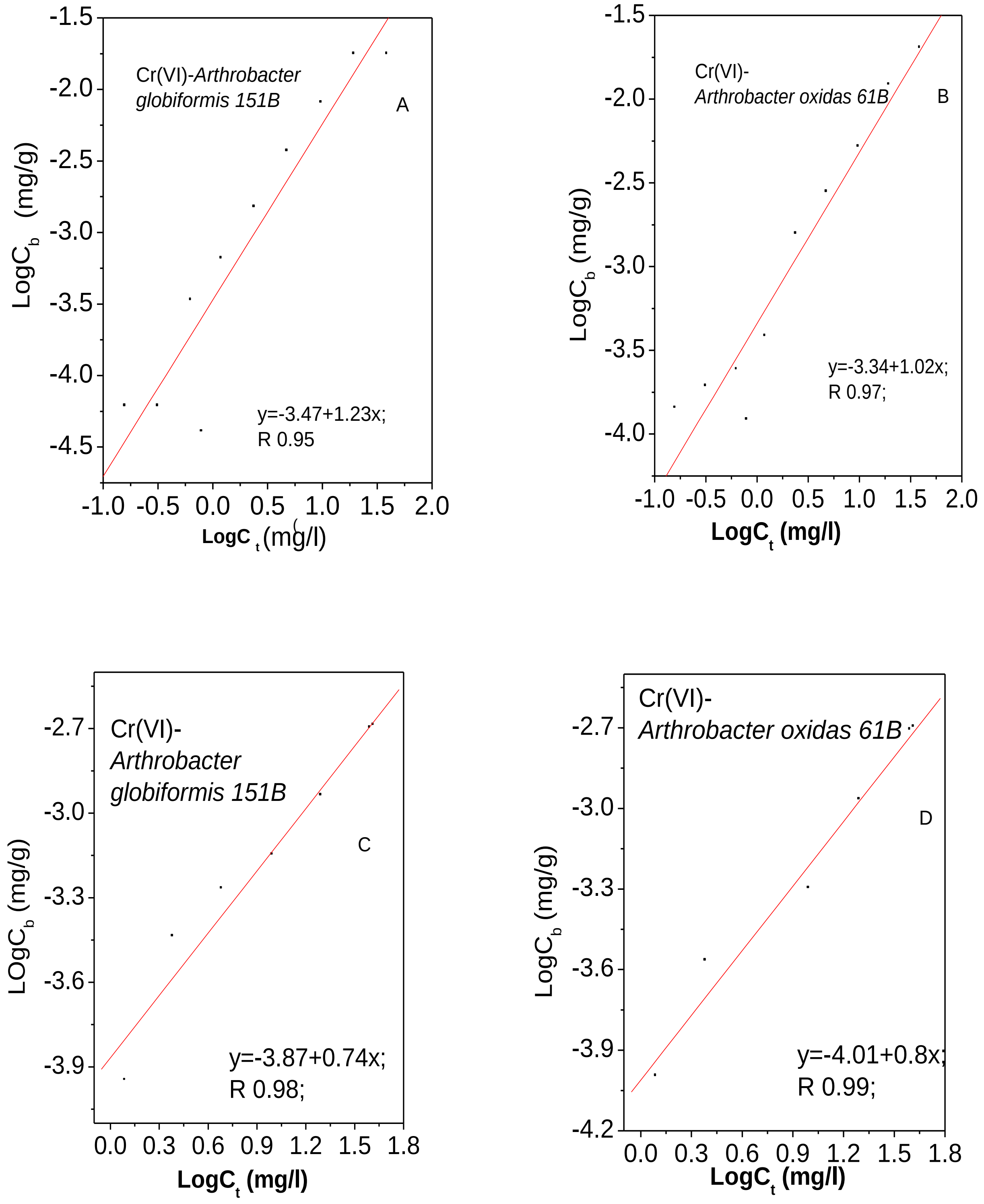

Fig.1 The linearized Freundlich adsorption isotherms of Cr (VI) ion-*Arthrobacter globiformis* and *Arthrobacter oxidas*. ($C_b$ is the binding metal concentration (mg/g) and $C_{total}$ is initial Cr concentration(mg/l). (A and B dry cells–, C and D living cells)

Table 1. Biosorption characteristics for Cr(VI)- and Cr(III) - *Arthrobater Oxidas* and Cr(VI)- and Cr(III) - *Arthrobacter Globiformis* at $23^0$ C

| | **Cr(VI)** | | | Cr(III) | | |
|---|---|---|---|---|---|---|
| .Biosorption characteristics (K, n) | *K* x $10^{-4}$ | *n* | $R^2$ | *K* x $10^{-4}$ | *n* | $R^2$ |
| *Arthrobater oxidas* (dry cells) | 4.6 | 1.25 | 0.98 | 26.0 | 1.37 | 0.98 |
| *Arthrobacter globiformis* (dry cells) | 3.4 | 1.35 | 0.96 | 20.2 | 1.23 | 0.98 |
| *Arthrobater oxidas* (living cells) | 1.0 | 1.25 | 0.94 | - | - | - |
| *Arthrobacter globiformis*(living cells) | 1.3 | 1.35 | 0.91 | - | - | - |
| *Arthrobater oxidas* + Zn(II) | 6.6 | 1.08 | 0.98 | - | - | - |
| *Arthrobacter globiformis*+Zn(II) | 8.1 | 1.19 | 0.96 | - | - | - |
| *Arthrobater oxidas* + Mn(II) | - | - | - | 24.3 | 1.41 | 0.96 |
| *Arthrobacter globiformis*+Mn(II) | - | - | - | 19.4 | 1.47 | 0.94 |

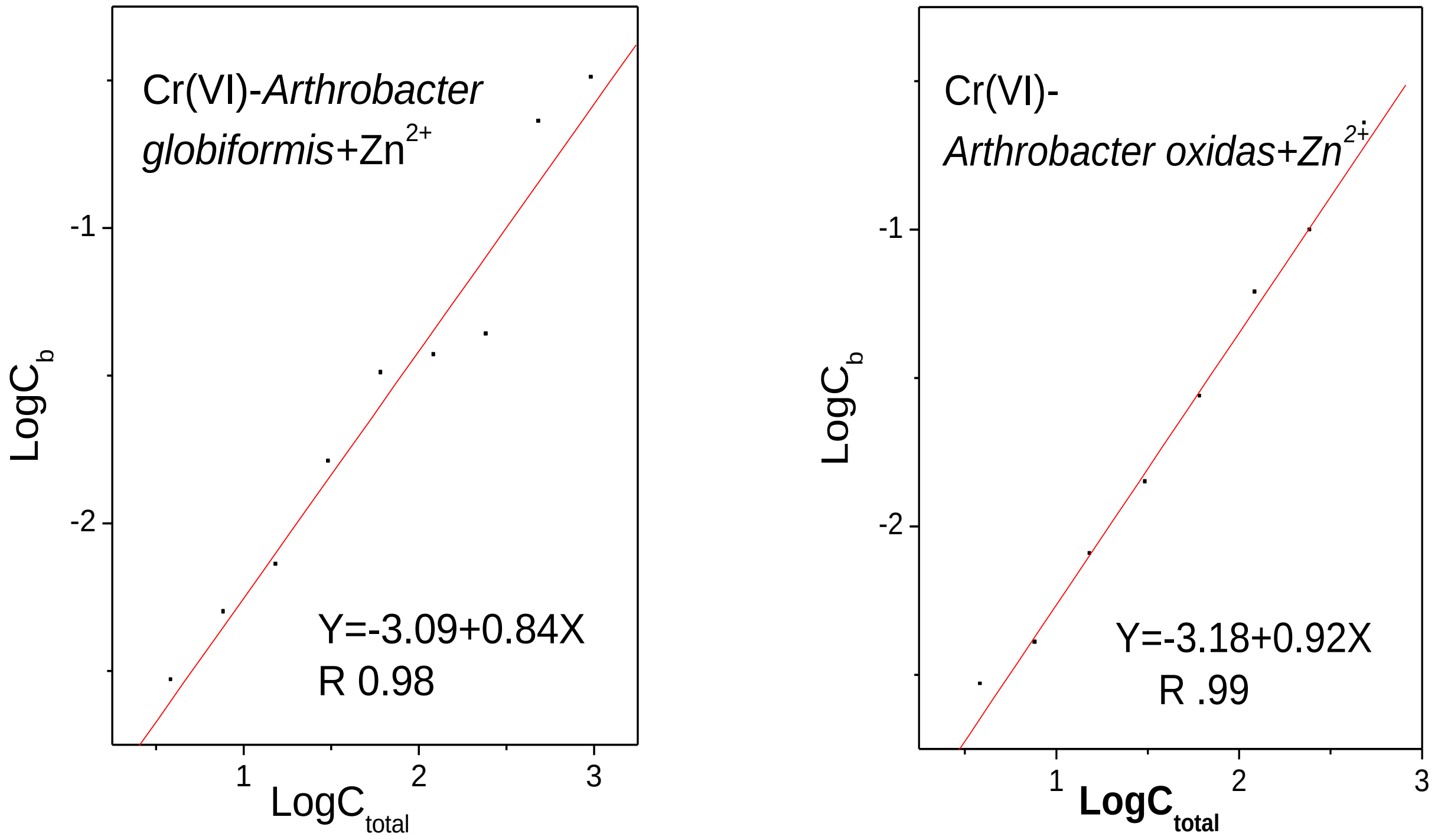

Fig.2. The linearized Freundlich adsorption isotherms of Cr (VI) ion-*Arthrobacter globiformis*+Zn and *Arthrobacter oxidas*+Zn. (The parameters are the same as in fig.1).

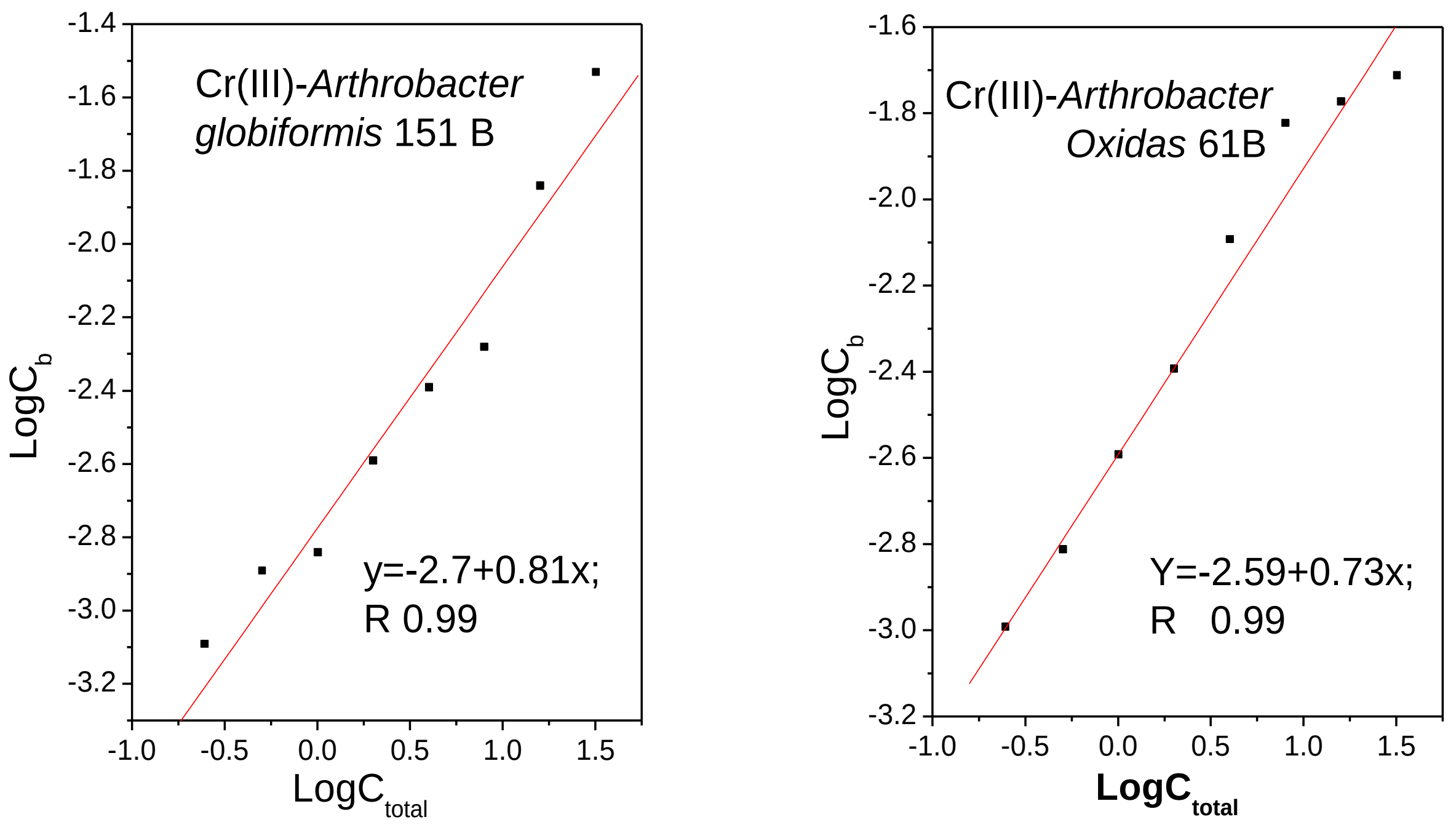

Fig.3. The linearized Freundlich adsorption isotherms of Cr (III) ion-*Arthrobacter globiformis* and *Arthrobacter oxidas* . (The parameters are the same as in fig.1).

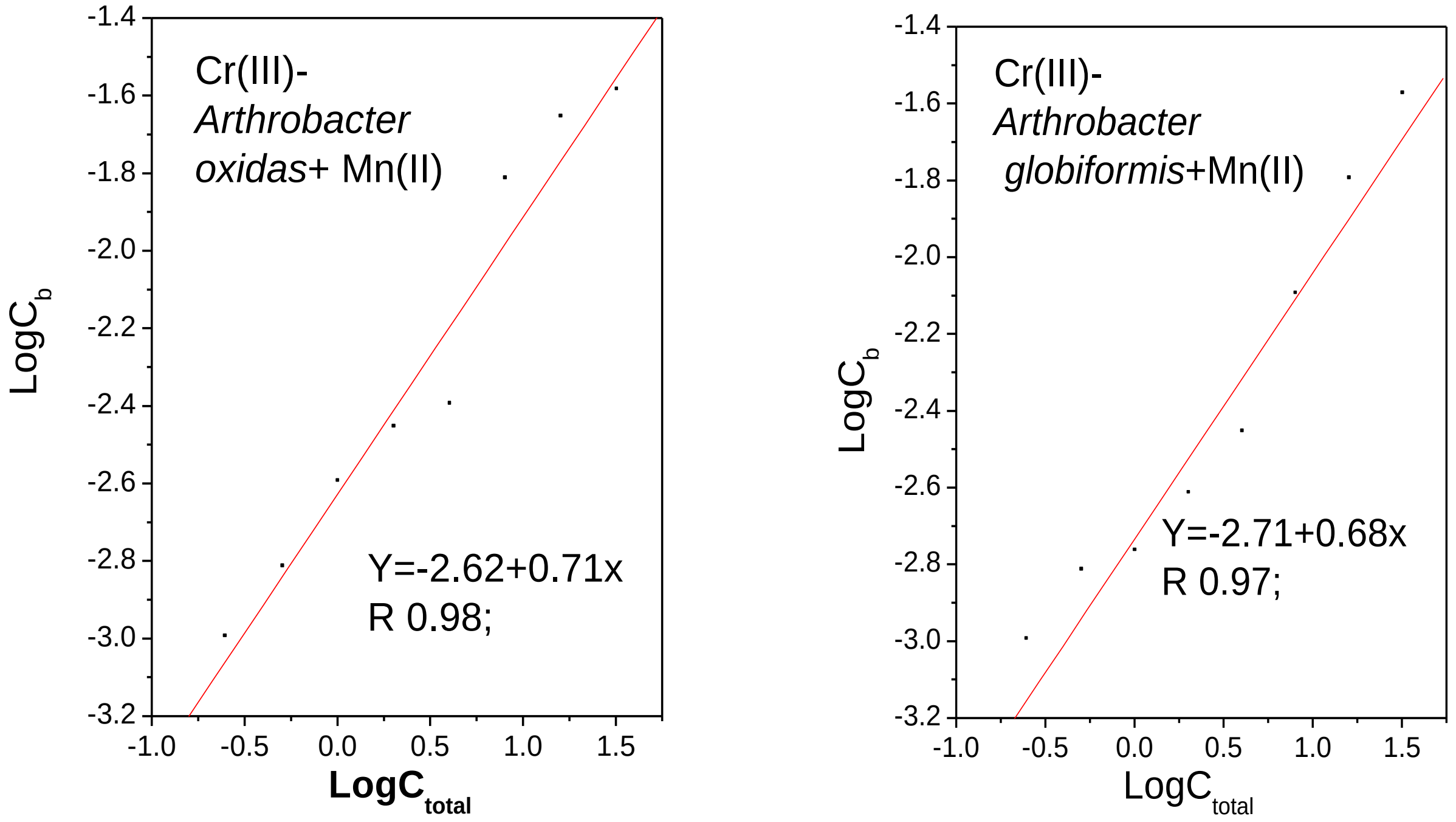

Fig.4. The linearized Freundlich adsorption isotherms of Cr (III) ion-*Arthrobactern globiformis*+Mn and *Arthrobacter oxidas*+Mn (The parameters are the same as in fig.1).